\documentclass[aps,preprintnumbers,amsmath,amssymb,showpacs,nofootinbib,twocolumn,superscriptaddress]{revtex4}
\usepackage{amsfonts}
\usepackage{mathrsfs}
\usepackage{graphicx}
\usepackage{amsmath}
\usepackage{amssymb}
\usepackage{bm}
\usepackage{bbm}
\usepackage{epsfig}
\usepackage{multirow}
\usepackage{float}
\usepackage{array}
\usepackage{color}
\usepackage{upgreek}
\usepackage{lipsum}
\makeatletter
\def\hlinew#1{%
  \noalign{\ifnum0=`}\fi\hrule \@height #1 \futurelet
   \reserved@a\@xhline}
\usepackage{array}
\newcommand{\PreserveBackslash}[1]{\let\temp=\\#1\let\\=\temp}
\newcolumntype{C}[1]{>{\PreserveBackslash\centering}p{#1}}
\newcolumntype{R}[1]{>{\PreserveBackslash\raggedleft}p{#1}}
\newcolumntype{L}[1]{>{\PreserveB    ackslash\raggedright}p{#1}}

\def\bfpsi{\mbox{\boldmath $\psi$}}
\def\bfeta{\mbox{\boldmath $\eta$}}

\newcommand{\nn}{\nonumber}
\newcommand{\bqa}{\begin{eqnarray}}
\newcommand{\eqa}{\end{eqnarray}}

\begin{document}

\title{\mbox{}\\[10pt]
Next-to-next-to-leading-order QCD corrections to $\bm{e}^{\bm +} \bm{e}^{\bm
-}\bm{\to}\bm{J}\bm{/}\bfpsi\bm{+}\bfeta_{\bm{c}}$ at $\bm{B}$
factories
}

\author{Feng Feng\footnote{F.Feng@outlook.com}}
\affiliation{Institute of High Energy Physics, Chinese Academy of
Sciences, Beijing 100049, China\vspace{0.2cm}}
\affiliation{China University of Mining and Technology, Beijing 100083, China\vspace{0.2cm}}

\author{Yu Jia\footnote{jiay@ihep.ac.cn}}
\affiliation{Institute of High Energy Physics, Chinese Academy of
Sciences, Beijing 100049, China\vspace{0.2cm}}
\affiliation{School of Physics, University of Chinese Academy of Sciences,
Beijing 100049, China\vspace{0.2cm}}

\author{Zhewen Mo\footnote{mozw@ihep.ac.cn}}
\affiliation{Institute of High Energy Physics, Chinese Academy of
Sciences, Beijing 100049, China\vspace{0.2cm}}
\affiliation{School of Physics, University of Chinese Academy of Sciences,
Beijing 100049, China\vspace{0.2cm}}

\author{Wen-Long Sang\footnote{wlsang@ihep.ac.cn}}
 \affiliation{School of Physical Science and Technology, Southwest University, Chongqing 400700, China\vspace{0.2cm}}

 \author{Jia-Yue Zhang\footnote{zhangjiayue@ihep.ac.cn}}
 \affiliation{Institute of High Energy Physics, Chinese Academy of
 Sciences, Beijing 100049, China\vspace{0.2cm}}
 \affiliation{School of Physics, University of Chinese Academy of Sciences,
 Beijing 100049, China\vspace{0.2cm}}

\date{\today}

\begin{abstract}
Within the nonrelativistic QCD (NRQCD) factorization framework,
we compute the long-awaited ${\mathcal O}(\alpha_s^2)$ correction for the
exclusive double charmonium production process at $B$ factories, {\it i.e.},
$e^+e^-\to J/\psi+\eta_c$ at $\sqrt{s}=10.58$ GeV. For the first time,
we confirm that NRQCD factorization does hold at next-to-next-to-leading-order (NNLO)
for exclusive double charmonium production.
It is found that including the NNLO QCD correction considerably reduces the
renormalization scale dependence, and also implies the
reasonable perturbative convergence behavior for this process.
Our state-of-the-art prediction is consistent with the
\textsc{BaBar} measurement within errors.
\end{abstract}

\maketitle

\noindent{\color{blue} \it 1.~Introduction.}
Back in the beginning of this century,
one particularly pressing dilemma of Standard Model is the
severe discrepancy between the \textsc{Belle} measurement~\cite{Abe:2002rb} and the subsequent predictions~\cite{Braaten:2002fi,Liu:2002wq,Hagiwara:2003cw} for the
exclusive double-charmonium production $e^+e^-\to J/\psi+\eta_c$ at $\sqrt{s}=10.58$ GeV.
This disquieting discrepancy has triggered a flurry of theoretical explorations in
the following years.
Although some explanations invoke certain exotic scenarios~\cite{Brodsky:2003hv,Cheung:2003xw},
the consensus is that this puzzle is rooted in our inadequate knowledge about
quarkonium production mechanism.
The mainstream investigations from the first principles of QCD are based on the light-cone factorization~\cite{Ma:2004qf,Bondar:2004sv,Braguta:2005kr,Bodwin:2006dm} and
NRQCD factorization~\cite{Zhang:2005cha,Gong:2007db,He:2007te,Bodwin:2007ga}.
Unfortunately, apart from poorly known light-cone distribution amplitudes of charmonia,
some unsurmountable difficulty in the former approach, {\it e.g.}, the endpoint singularity,
renders a next-to-leading order (NLO) perturbative calculation to such a helicity-flipped
exclusive process impossible~\cite{Chernyak:1983ej}.
In contrast, for the hard exclusive process $e^+e^-\to J/\psi+\eta_c$,
the NRQCD approach~\cite{Bodwin:1994jh} provides a more predictive framework that
is amenable to systematically incorporating the higher-order perturbative and relativistic corrections.

One key progress in alleviating the tension is brought by the NLO perturbative
calculation for $e^+e^-\to J/\psi+\eta_c$ in NRQCD approach,
where a significant positive ${\mathcal O}(\alpha_s)$ correction is found~\cite{Zhang:2005cha,Gong:2007db}.
The relative ${\mathcal O}(v^2)$ correction to $e^+e^-\to
J/\psi+\eta_c$ has also been addressed~\cite{Braaten:2002fi,He:2007te,Bodwin:2007ga},
where $v$ denotes the typical velocity of the $c$ quark inside a charmonium.
Notwithstanding large uncertainty inherent to various NRQCD matrix elements,
it was suggested that~\cite{He:2007te,Bodwin:2007ga},
by including both ${\cal O}(\alpha_s)$ and (a partial resummation of)
relativistic corrections, one may largely resolve the discrepancy.
Later the joint perturbative and relativistic order-$\alpha_s v^2$ correction was
also investigated, which was found to modestly enhance the existing NRQCD predictions~\cite{Dong:2012xx,Xi-Huai:2014iaa}.

The recently commissioning \textsf{Belle II} experiment will certainly conduct
more precise measurement for this double quarkonium production channel.
Therefore, it is desirable to have more accurate theoretical prediction available.
Given the substantial ${\cal O}(\alpha_s)$ correction to the cross section,
one cannot resist speculating whether the magnitude of the NNLO perturbative correction
is abnormally large or not.
Would the ${\cal O}(\alpha_s^2)$ correction for $e^+e^-\to J/\psi+\eta_c$
be as significant as the recently available NNLO perturbative corrections for
$\gamma^*\gamma\to \eta_c$~\cite{Feng:2015uha} and $\eta_c\to {\rm light\;hadrons}$~\cite{Feng:2017hlu}?
Undoubtedly, the NNLO correction for a $1\to 4$ process
involving massive quarks represents a cutting-edge challenge
in the area of multi-loop calculation.
To sense the daunting difficulty,
we quote the authoritative review of quarkonium physics in 2011~\cite{Brambilla:2010cs}:
{\it ``the calculation of \ldots is perhaps beyond the current state of the art''}.
Notwithstanding enormous technical obstacles, in this paper we will report our endeavour
in accomplishing this NNLO calculation.

\vspace{0.2 cm}
\noindent{\color{blue} \it 2.~NRQCD factorization for cross section.}
It is convenient to define the time-like electromagnetic (EM) form factor $F(s)$
through
\begin{align}
\langle J/\psi(P_1,\lambda)+\eta_c(P_2)\vert J_{\text EM}^\mu \vert 0
\rangle = i\,F (s)\,\epsilon^{\mu\nu\rho\sigma} P_{1\nu} P_{2 \rho}
\varepsilon^*_\sigma (\lambda),
\nn\\
\label{EMFM:jpsi+etac}
\end{align}
where $J^\mu_{\text EM}$ is the quark EM current, and $s=(P_1+P_2)^2$.
The tensor structure specified in (\ref{EMFM:jpsi+etac}) is uniquely constrained
by Lorentz and parity invariance. The outgoing $J/\psi$ must be transversely polarized,
{\it i.e.}, $\lambda=\pm 1$.

For hard exclusive reaction involving quarkonium, NRQCD factorization also holds at amplitude level.
Specifically speaking, the EM form factor in (\ref{EMFM:jpsi+etac}) can be expressed as
\begin{align}
\label{EMFF:NRQCD:factorization}
F(s) =&  \sqrt{4 M_{J/\psi} M_{\eta_c}} \langle J/\psi| \psi^\dagger
\bm{\sigma} \cdot \bm{\epsilon} \chi |0\rangle \langle \eta_c|
\psi^\dagger \chi | 0 \rangle \nn\\
&\times\left[f + g_{J/\psi} \langle v^2 \rangle_{J/\psi} + g_{\eta_c} \langle
v^2 \rangle_{\eta_c} + \cdots\right],
\end{align}
where the perturbatively calculable effects are encoded in the
short-distance coefficients (SDCs) $f$ and $g_H\,(H=J/\psi, \eta_c)$,
and the long-distance effects encapsulated in the nonperturbative
vacuum-to-charmonium matrix elements, which are often modeled by
the phenomenological charmonium wave functions at the origin.
Note by default the charmonium states in the NRQCD matrix elements are
at rest and nonrelativistically normalized. The prefactor in
\eqref{EMFF:NRQCD:factorization} compensates the fact that these states
are relativistically normalized in \eqref{EMFM:jpsi+etac}. $\langle v^2 \rangle_{J/\psi}$ and $\langle v^2 \rangle_{\eta_c}$ are defined as the dimensionless ratios of two NRQCD matrix elements for $J/\psi$ and $\eta_c$~\cite{Braaten:2002fi}, which characterize the size of relativistic correction.

Substituting~\eqref{EMFF:NRQCD:factorization} into (\ref{EMFM:jpsi+etac}), it is straightforward to deduce the cross section, which can be further divided into the ${\cal O}(v^0)$ and ${\cal O}(v^2)$ pieces:
\begin{align}
\label{cross:section:decomposition}
& \sigma[e^+e^-\to J/\psi + \eta_c] = \dfrac{4\pi \alpha^2}{3}
\left(\dfrac{|{\bf P}|}{\sqrt{s}}\right)^3 \left|F(s)\right|^2
\nn\\
&=  \sigma_0+\sigma_2+{\mathcal O}(\sigma_0 v^4),
\end{align}
where $|{\bf P}|$ signifies the magnitude of the three-momentum carried
by the $J/\psi$ in the center-of-mass frame, and
\begin{subequations}
\begin{align}
\sigma_0  =&  \dfrac{8\pi\alpha^2 m^2 (1-4 r)^{3/2}}{3}\langle
{\mathcal O} \rangle_{J/\psi} \langle{\mathcal
O}\rangle_{\eta_c} \,|f|^2,
\label{sigma_0:definition}
\\
\sigma_2 =&  \dfrac{4\pi \alpha^2 m^2 (1-4 r)^{3/2}}{3} \langle
{\mathcal O} \rangle_{J/\psi} \langle{\mathcal
O}\rangle_{\eta_c}
\label{sigma_2:definition}
\\
&\times
\sum_{H=J/\psi,\eta_c}\bigg(\dfrac{1-10r}{1-4r} |f|^2
 + 4\,{\rm Re}(f g_{H}^*)
\bigg) \langle v^2\rangle_H,
\nn
\end{align}
\label{sigma0:sigma2:definitions}
\end{subequations}
where a dimensionless ratio
$$
r={4 m^2/s}
$$
is introduced for brevity.
To condense the notation, we have also introduced the following symbols:
$\langle \mathcal{O}\rangle_{J/\psi}=\big| \langle
J/\psi|\psi^\dag \bm{\sigma} \cdot \bm{\epsilon} \chi |0\rangle
\big|^2$, and $\langle {\mathcal O} \rangle_{\eta_c} = \big|
\langle {\eta_c}| \psi^\dag \chi |0\rangle \big|^2$.
In deriving (\ref{sigma0:sigma2:definitions}), we have employed the
Gremm-Kapustin relation~\cite{Gremm:1997dq} $M_H^2\approx
4m^2(1+\langle v^2 \rangle_H)$ to eliminate the explicit
occurrence of the charmonium masses.

Thanks to the weaker strong coupling constant $\alpha_s$ at the scale of
the charm quark Compton wavelength or shorter,
the SDCs $f$ and $g_{H}$ are subject to perturbative expansion in $\alpha_s$:
\begin{subequations}
\begin{align}
f =& f^{(0)} + \dfrac{\alpha_s}{\pi} f^{(1)}  + \dfrac{\alpha_s^2}{\pi^2} f^{(2)}  + \cdots,
\\
g_{H} =& g^{(0)}_{H} + \dfrac{\alpha_s}{\pi}  g^{(1)}_{H}  + \cdots.
\end{align}
\label{Expanding:SDCs:f:g}
\end{subequations}

Substituting \eqref{Expanding:SDCs:f:g} back to \eqref{sigma0:sigma2:definitions},
we can organize $\sigma_0$ and $\sigma_2$ in perturbation series in $\alpha_s$.
For example, using
\begin{align}
|f|^2  =& \big|f^{(0)}\big|^2+ \dfrac{\alpha_s}{\pi}2{\rm Re} \left(f^{(0)}f^{(1)*} \right)
\\
& + \left(\dfrac{\alpha_s}{\pi}\right)^2 \left[2 {\rm Re} \left(f^{(0)}f^{(2)*}\right) + \big|f^{(1)}\big|^2\right],
\nn
\end{align}
we can decompose
$\sigma_0= \sigma^{(0)}_0+\sigma^{(\alpha_s)}_0+\sigma^{(\alpha_s^2)}_0+\cdots$.

The tree-level SDCs through ${\mathcal O}(v^2)$ have been
available long ago~\cite{Braaten:2002fi}. Here we list their values:
\begin{subequations}
\begin{align}
& f^{(0)} = \dfrac{32 \pi C_F e_c \alpha_s }{N_c\,m s^2},
\\
& g_{J/\psi}^{(0)} = \dfrac{3-10r}{6}  f^{(0)},
\quad
g_{\eta_c}^{(0)} = \dfrac{2-5r}{3}f^{(0)},
\end{align}
\label{c_i:tree-level:values}
\end{subequations}
where $e_c={2/3}$ is the electric charge of the charm quark,
$N_c=3$ is the number of colors and $C_F=(N_c^2-1)/(2N_c)$.

The NLO perturbative corrections to those SDCs, for both $f$~\cite{Zhang:2005cha,Gong:2007db} and $g_H$~\cite{Dong:2012xx,Xi-Huai:2014iaa},
have also been known for a while for.
Since their analytic expressions are rather lengthy, here we just list
the asymptotic expression for $f^{(1)}$ as $\sqrt{s} \gg m$:
\bqa
&& f^{(1)}\vert_{\text{asym}} =
f^{(0)}  \Bigg\{ \beta_0 \bigg(- \frac{1}{4}
\ln\frac{s}{4\mu_R^2} + \frac{5}{12}\bigg) + \bigg( \frac{13}{24}
\ln^2 r
\nn\\
&&\qquad + \frac{5}{4} \ln2 \ln r - \frac{41}{24} \ln r -
\frac{53}{24}\ln^22 +\frac{65}{8}\ln2 -\frac{1}{36}\pi^2
\nn\\
&&\qquad -\frac{19}{4} \bigg)
 + i\pi \bigg( \frac{1}{4}\beta_0 + \frac{13}{12}\ln r  +
\frac{5}{4} \ln2- \frac{41}{24} \bigg)  \Bigg\},
\label{c_0:NLO:asym:expressions}
\eqa
where $\beta_0={11\over 3} C_A-{2\over 3}n_f$ is the one-loop
coefficient of the QCD $\beta$ function, and $n_f=4$ denotes the
number of active quark flavors. $\mu_R$ denotes the renormalization
scale, with the natural choice around $\sqrt{s}/2$. The double logarithm $\alpha_s \ln^2 r$ is believed to account for large
positive NLO correction, which was first discovered in~\cite{Jia:2010fw}, and carefully analyzed in~\cite{Bodwin:2014dqa}. The asymptotic expressions for $g_{J/\psi}^{(1)}$ and $g_{\eta_c}^{(1)}$ can be found in \cite{Dong:2012xx}.

\begin{figure}[tbh]
\begin{center}
\includegraphics[width=0.47\textwidth]{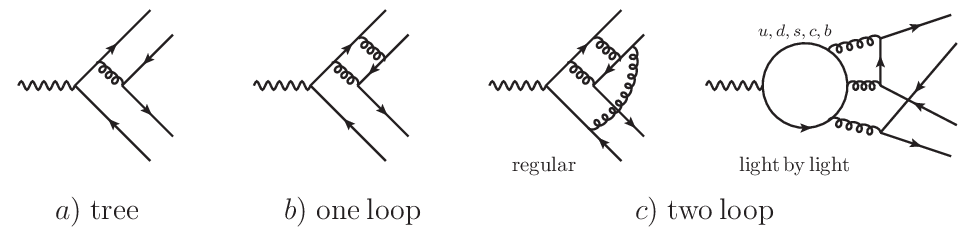}
\caption{Representative diagrams for $\gamma^*\to c\bar{c}({}^3S_1^{(1)})+c\bar{c}({}^1S_0^{(1)})$
through NNLO in $\alpha_s$.
\label{feynman:diagrams}}
\end{center}
\end{figure}

\vspace{0.2 cm}
\noindent{\color{blue} \it 3.~Outline of calculation and main result.}
To compute $f^{(2)}$, we take the shortcut by directly calculating the quark-level amplitude for
$\gamma^*\to c\bar{c}({}^3S_1^{(1)},P_1)+c\bar{c}({}^1S_0^{(1)},P_2)$.
To LO accuracy in $v$, we neglect the relative momentum in each $c\bar{c}$ pair prior to carrying out the
loop integration, which amounts to directly extracting the NRQCD SDCs from the hard region.
We work in $d=4-2\epsilon$ spacetime dimensions to regularize both UV and IR divergences.
About $2000$ NNLO Feynman diagrams, as well as the corresponding amplitudes are generated
by {\tt QGraf}/{\tt FeynArts}~\cite{Nogueira:1991ex,Hahn:2000kx}.
Some representative diagrams through NNLO are sampled in Fig.~\ref{feynman:diagrams}.
Since the center-of-mass energy at $B$ factory exceeds twice bottom quark mass, we explicitly include the bottom loops in both regular and light-by-light diagrams.
It is legitimate to drop those ``light-by-light'' diagrams in which $J^\mu_\text{EM}$ directly couples
with the light quark, since $\sum_{q=u,d,s} e_q=0$.
The covariant projector technique is utilized to project each $c\bar{c}$ pair onto the intended quantum number.
We then employ the packages {\tt FeynCalc}/{\tt FormLink}~\cite{Mertig:1990an,Feng:2012tk} to
conduct the trace over Dirac and SU$(N)$ color matrices.
After the integration-by-parts (IBP) reduction with the aid of {\tt Apart}~\cite{Feng:2012iq}
and {\tt FIRE}~\cite{Smirnov:2014hma}, we end up with about $700$ master integrals (MIs). Originally, we first tried to employ the sector decomposition (SD) method to evaluate these MIs numerically and find it extremely challenging to obtain reliable results within tolerable amount of time.
Fortunately,  a powerful new algorithm dubbed Auxiliary Mass Flow
(AMF) has recently been developed by Liu and Ma~\cite{Liu:2017jxz,Liu:2020kpc}. This algorithm is based on numerical
differential equation method, which is tailored to tackle multi-scale multi-loop MIs with high numerical precision in a short time.
We utilize the recently released package {\tt AMFlow}~\cite{Liu:2022chg} to compute all the MIs.

To eliminate UV divergences, we employ the field-strength and mass renormalization,
with two-loop expressions of $Z_2$ and $Z_m$ taken from \cite{Broadhurst:1991fy, Melnikov:2000zc,Baernreuther:2013caa},
and renormalize the strong coupling constant in the $\overline{\rm MS}$ scheme to two-loop order.
However, the renormalized NNLO QCD amplitude is found to still contain a single IR pole,
yet with the coefficient exactly equal to the sum of the anomalous dimensions
for the NRQCD bilinear operators
carrying the quantum number of $J/\psi$~\cite{Czarnecki:1997vz,Beneke:1997jm} and $\eta_c$~\cite{Czarnecki:2001zc}.
This pattern is exactly what we anticipate for NRQCD factorization for double quarkonium production at NNLO.
This IR pole can be factored into the corresponding NRQCD matrix elements
under the $\overline{\rm MS}$ scheme, which are actually scale-dependent quantities whose evolution is given by the following  equation:
\begin{subequations}
\begin{align}
&\dfrac{\mathrm{d} \ln\langle J/\psi| \psi^\dagger
\bm{\sigma} \cdot \bm{\epsilon} \chi |0\rangle }{\mathrm{d}\ln\mu_\Lambda^2}=-\left(\dfrac{\alpha_s}{\pi}\right)^2\gamma_{J/\psi}+\mathcal{O}\left(\alpha_s^3,v^2\right),
\\
&\dfrac{\mathrm{d} \ln\langle \eta_c|
\psi^\dagger \chi | 0 \rangle}{\mathrm{d}\ln\mu_\Lambda^2}=-\left(\dfrac{\alpha_s}{\pi}\right)^2\gamma_{\eta_c}+\mathcal{O}\left(\alpha_s^3,v^2\right),
\end{align}\label{LDME:RG eq}
\end{subequations}
with
\begin{subequations}
\begin{align}
\gamma_{J/\psi} &= -\frac{\pi^2}{12} C_F \left( 2C_F + 3C_A \right),
\\
\gamma_{\eta_c} &= -\frac{\pi^2}{4} C_F \left( 2C_F + C_A \right),
\end{align}
\end{subequations}
where $\mu_\Lambda$ is referred to as the NRQCD factorization scale, whose value lies somewhere between $m v$ and $m$.
Finally, the UV, IR-finite ${\cal O}(\alpha_s^2)$ SDC reads
\begin{align}
f^{(2)} =& f^{(0)} \bigg\{ \dfrac{\beta_0^2}{16} \ln^2\frac{s}{4\mu_R^2} -
\left( \frac{\beta_1}{16} + \frac{1}{2} \beta_0 \hat{f}^{(1)}\right) \ln\frac{s}{4\mu_R^2}
\nn\\
&+(\gamma_{J/\psi}+\gamma_{\eta_c})\ln\frac{\mu_\Lambda^2}{m^2} + \textsf{F}(r) \bigg\}.
\label{f2:expressions}
\end{align}

Moreover, $\hat{f}^{(1)} \equiv f^{(1)}/f^{(0)}\big\vert_{\mu_R=\sqrt{s}/2}$,
$\beta_1 = \frac{34}{3}C_A^2 -\frac{20}{3}C_A T_F n_f - 4C_F T_F n_f$ is the two-loop coefficient of
the QCD $\beta$ function. The occurrence of $\ln \!\mu_R$
is dictated by the renormalization group invariance.

The non-logarithmic term is embedded in the function $\textsf{F}(r)$ in \eqref{f2:expressions}.
It is of our primary interest to ascertain this term as precise as possible, in order to
pin down the impact of the NNLO perturbative correction.

\vspace{0.2 cm}
\noindent{\color{blue} \it 4.~Phenomenology.}
The production rate initially measured by \textsc{Belle} is
$\sigma[e^+e^-\to J/\psi+\eta_c]\times{\mathcal B}_{\ge 4}=33_{-6}^{+7}\pm 9$
fb~\cite{Abe:2002rb}, later shifted to $\sigma[J/\psi+\eta_c] \times
{\mathcal B}_{>2} = 25.6\pm 2.8\pm3.4$ fb~\cite{Abe:2004ww}, where
${\mathcal B}_{>n}$ denotes the branching fraction for the $\eta_c$
into $n$ charged tracks. An independent measurement by \textsc{BaBar}
in 2005 yields $\sigma[J/\psi+\eta_c] \times
{\mathcal B}_{>2}=17.6\pm2.8^{+1.5}_{-2.1}$ fb~\cite{Aubert:2005tj}.

In the numerical analysis, We concentrate on the $B$ factories center-of-mass
energy $\sqrt{s}=10.58$ GeV, and take charm quark pole mass $m=1.5\,\mathrm{GeV}$ and bottom mass $m_b=4.7\,\mathrm{GeV}$.
The QED coupling constant $\alpha(\sqrt{s}) = 1/130.9$ \cite{Bodwin:2007ga}, and the QCD running coupling constant is evaluated to two-loop accuracy
with the aid of the package {\tt RunDec}~\cite{Chetyrkin:2000yt}.
The NRQCD matrix elements are taken from \cite{Bodwin:2007ga}: $\langle {\cal O} \rangle_{J/\psi} = 0.440\:{\rm GeV}^3$ and
$\langle {\cal O} \rangle_{\eta_c} = 0.437\:{\rm GeV}^3$. For simplicity, we omit the relativistic corrections in phenomenological analysis.

\begin{figure}[tbh]
\begin{center}
\includegraphics[width=0.45\textwidth]{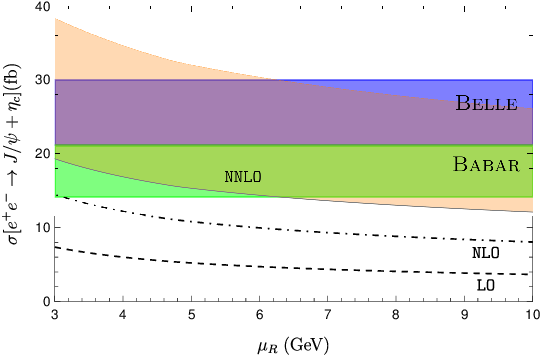}
\caption{The cross section, predicted with various level of precision, as function of $\mu_R$.
We take $m=1.5$ GeV. The brown bands represent the uncertainty due to varying $\mu_\Lambda$ from $1\,\mathrm{GeV}$ to $m$, where the lower bound corresponds to $\mu_\Lambda=1\,\mathrm{GeV}$ and upper bound $\mu_\Lambda=m$.
\label{X:sections:various:level}}
\end{center}
\end{figure}

In Fig.~\ref{X:sections:various:level}, we plot the dependence of the predicted cross section on $\mu_R$ and $\mu_\Lambda$,
including numerous individual contribution from different perturbative order. We  observe flatter $\mu_R$ dependence of the NNLO cross section.

In Table~\ref{Table:X:Section:component}, we enumerate the individual contribution to the
cross section at various levels of perturbative accuracy in NRQCD factorization.

\renewcommand\arraystretch{1.5}
\begin{table*}
\caption{Individual contributions to the predicted $\sigma[e^+e^-\to
  J/\psi+\eta_{c}]$ (in units of fb) at $\sqrt{s}=10.58$ GeV. We take
  $\mu_R=\sqrt{s}/2$, and $\mu_\Lambda = 1\,\mathrm{GeV}$. The first error is obtained by varying $m_c$ from $=1.3$ to $1.7\,\mathrm{GeV}$,
  and the second error is deduced by varying $\mu_R$ from $2m_c$ to $\sqrt{s}$.\label{Table:X:Section:component}}
  \begin{tabular}{|ccccc|}
    \hline
    $m(\mathrm{GeV})$ & $\mu_\text{R}$&{\tt LO}
    & {\tt NLO}
    & {\tt NNLO}
    \\
    \hline
    $1.5$ & $\sqrt{s}/2$
    &$5.05^{+0.92}_{-0.99}{}^{+2.31}_{-1.49}$
    &$10.54^{+2.86}_{-2.60}{}^{+3.92}_{-2.66}$
    &$15.00^{+5.03}_{-4.14}{}^{+4.29}_{-3.14}$
    \\
    \hline
    \end{tabular}
  \end{table*}

From Table~\ref{Table:X:Section:component}, we observe that
the NNLO correction to this double charmonium production process is sizable,
but not yet as substantial as the NLO correction (Note this is not the case in \cite{Feng:2015uha,Feng:2017hlu}).
It is reassuring that the perturbative expansion exhibits a convergent signature.
Depending on the choice of $\mu_R$, the NNLO correction may range from $33\%$ to $51\%$.
Nevertheless, the total NNLO cross section possesses a milder dependence on the renormalization scale
than the NLO prediction. We note that a recent study~\cite{Sun:2018rgx} based on principle of maximum conformality
claims a favored setting of $\mu_R$ to be around $2-3$ GeV, which can bring NLO NRQCD prediction
consistent with the $B$ factory measurements.

In Table~\ref{Table:X:Section:component} we also include the dependence of the double charmonium production cross section on charm quark mass.
Table~\ref{Table:X:Section:component} indicates that the NLO and NNLO NRQCD predictions are quite sensitive to the charm pole mass,
and smaller mass yields a prediction much closer to the data.

\begin{table*}[htb]
    \caption{Values of $\textsf{F}(r)$ at some different center-of-mass energies with $m=1.5\,\mathrm{GeV}$.
    \hfill\hfill \label{Table:1}}
    \begin{tabular}{|c|c|}
    \hline
    $\sqrt{s} (\mathrm{GeV})$ & $\textsf{F}(r)$\\
    \hline
    $10.58$ & $(25.300-19.883 i)-(0.18765+0.01218 i)_\text{lbl}$ \\
    $91.2$  & $(589.13-308.17 i)-(0.012676+0.010692  i)_\text{lbl}$ \\
    $240$ %
    & $(1178.68-556.97i)-(0.0092905+0.0013069i)_\text{lbl}$ \\
     $350$ 
     & $(1490.9-678.7i)-(0.0096519-0.0017494i)_\text{lbl}$ \\
     $500$ 
    & $(1835.9-807.6i)-(0.0105894-0.0044144i)_\text{lbl}$ \\
    \hline
    \end{tabular}
    \end{table*}
\begin{subequations}
\label{Value:Re:F:two:point}
\end{subequations}

To sense the profile of $\textsf{F}(r)$, we list the values of this function at some benchmark energy points in TABLE~\ref{Table:1}.
The terms labeled with subscript ``lbl" denote the contributions from the ``light-by-light" diagrams,
as illustrated by the representative diagram in Fig.~\ref{feynman:diagrams}.
The numerical difficulty to obtain accurate predictions increases enormously as $\sqrt{s}$ increases.
Our results can be applied to predict the exclusive $J/\psi+\eta_c$ production at future very high energy $e^+e^-$ colliders
such as {\texttt Z} factory, \texttt{CEPC}/\texttt{FCC-ee}, \texttt{ILC}, in which
the exclusive double charmonium production rates would be too small to be observed.
From the data of TABLE~\ref{Table:1}, we attempt to fit
the coefficient of the anticipated endpoint logarithm $\alpha_s^2\ln^4 r$. Pitifully,
perhaps because the maximum value of $\sqrt{s}$ ($500$ GeV) is still not asymptotically high,
we fail to determine this coefficient in an unambiguous manner.

\vspace{0.2 cm}
\noindent{\color{blue} \it 5.~Summary and outlook.}
More than one decade after the NLO correction became first available~\cite{Zhang:2005cha},
we eventually accomplish the long-awaited calculation of the ${\mathcal O}(\alpha_s^2)$
correction to $e^+e^-\to J/\psi+\eta_c$ at $B$ factories.
We verify that NRQCD factorization does hold at NNLO in $\alpha_s$ for exclusive
double $S$-wave charmonium production.
Including the NNLO QCD correction reduces the dependence on
the renormalization scale, and exhibits reasonable perturbative convergence behavior.
Our state-of-the-art prediction is compatible with the \textsc{BaBar} measurement, but still
somewhat smaller than the \textsc{Belle} measurement. The future remeasurement of this process
at \textsc{Belle II} experiment will be crucial to clarify the situation.
The future work along this direction includes precisely deducing the endpoint logarithm
$\propto \alpha_s^2\ln^4 r$ in ${\tt F}(r)$, and strive to resum these types of endpoint
logarithms to all orders.

\begin{acknowledgments}
The work of F. F. is supported by the National
Natural Science Foundation of China under Grant No. 11875318, No. 11505285, and by
the Yue Qi Young Scholar Project in CUMTB.
The work of Y.~J., Z.~M., and J.-Y.~Z. is supported in part by the National Natural Science Foundation of China under Grants No. 11925506, 11875263,
No. 11621131001 (CRC110 by DFG and NSFC).
The work of W.-L. S. is supported by the National Natural Science Foundation of China
under Grants No. 11975187.
\end{acknowledgments}

\end{document}